\documentclass[aps,pre,onecolumn,showpacs,showkeys,a4paper]{revtex4}
\input epsf  
\usepackage{graphicx} 
\usepackage{amsmath}
\usepackage{amssymb}
\usepackage{amscd}

\begin{document} 

 \title{Phase diagram of the random frequency oscillator: \\ 
   The case of  Ornstein-Uhlenbeck noise}
\author{Kirone Mallick and Pierre-Emmanuel Peyneau}
  \affiliation{Service de Physique Th\'eorique, Centre d'\'Etudes de Saclay,
 91191 Gif-sur-Yvette Cedex, France}
  \email{mallick@spht.saclay.cea.fr}
 \author{}
  \affiliation{}
  \email{}
%  \date{\today}

\begin{abstract}
   We study the stability of a stochastic
 oscillator whose frequency is a random process with finite
 time memory  represented  by an Ornstein-Uhlenbeck noise. 
 This system undergoes a noise-induced bifurcation when the
 amplitude of the noise grows. The critical curve,  
 that separates the absorbing phase from an extended
 non-equilibrium steady state,   corresponds to  
  the vanishing of the Lyapunov exponent that measures
 the asymptotic logarithmic  growth rate of the energy. We derive various
 expressions for 
 this Lyapunov exponent by using  different  approximation schemes. 
 This  allows  us to study
 quantitatively   the phase diagram of the random parametric  oscillator. 
  \end{abstract}
\maketitle

 \section{Introduction}

   Noise can modify   drastically the 
  the phase diagram of a dynamical 
  system   \cite{vankampen,gardiner,anishchenko}.
 Because of stochastic fluctuations of the control
 parameter, the critical value of the threshold  can  change and  noise
 can delay  or favor  a  phase transition \cite{lefever}. In the first  case,
 randomness can be useful as a stabilizing mean  and in  the second,
 noise may help   trigger  a phase transition that  is   otherwise
 very difficult to achieve;  for example, in  the dynamo
 effect, the role of the noise generated by   fluid turbulence
 is not  well understood at present   and it is  possible that 
  the critical  magnetic Reynolds number decreases 
 with noise \cite{fauve}. 
 In certain cases,   a physical system   subject to noise   undergoes 
  bifurcations  into states  that have no deterministic
  counterparts: the  stochastic phases  generated by randomness
  have   specific   characteristics
  (such as scaling behavior or  critical exponents) that   define 
  new universality classes \cite{munoz, pikovskybook}.   

  One of the simplest systems  that can be used as a paradigm for the
 study of noise-induced phase transitions  is the random frequency 
 oscillator \cite{luecke,ebeling}.  For instance, in practical 
    engineering problems, the Duffing oscillator
  with random frequency has been used as a model to study
 stability  of structures
 subject to random external forces, such as earthquakes, wind
 or ocean waves \cite{rong1,rong2,huang,xie}.  Whereas a
  deterministic oscillator with damping evolves 
 towards the unique  equilibrium  state of minimal energy,   the
  behavior  changes    if  
 the frequency of the oscillator  is  a time-dependent variable. 
   Due to continuous energy injection into the
 system through the frequency variations,  the system may sustain
 non-zero oscillations even in the long time limit.  The case when  the
 frequency is a periodic function of time 
  is the  classical problem  of parametric resonance  known as the 
 Mathieu oscillator; the  phase diagram   
 is  obtained by calculating the Floquet exponents  defined as  the
 characteristic growth rates of the amplitude of the system  \cite{nayfeh}.
  When the frequency of the pendulum is  a random process,
 the role of the Floquet exponents is taken over by the Lyapunov
 exponents \cite{arnold,pikovsky}.
 The system undergoes a bifurcation when the largest
 Lyapunov exponent,   defined as
  the  growth rate of the logarithm  of the  energy, changes its sign.
 Thus, the  Lyapunov  exponent vanishes on the 
  critical surface  that separates the phases in the parameter space.
  This criterion   involving  the sign of the  Lyapunov exponent
  has  a firm mathematical basis 
  and clarifies   the ambiguities that were found 
 in the study of  the stability of higher  moments \cite{bourret,arnold}.

    In a recent work \cite{philkir1},  we have carried out 
 an analytic  study of  the  phase diagram
 of the random oscillator driven by a  Gaussian white noise frequency. 
  We  have  shown \cite{philkir2} that,  
  in the case of an  inverted pendulum,  
  the unstable fixed point can be stabilized by noise and  
  a  noise-induced reentrant transition  occurs. 
 These results are   based on an exact  formula for the
 Lyapunov exponent \cite{hansel,tessieri,imkeller}.   
 In  the present  work,  we  intend to study  the phase diagram of an 
 oscillator whose frequency is a random process with finite
 time memory. More precisely, we  consider here the case of 
 an  Ornstein-Uhlenbeck noise of  correlation time $\tau$. 
 From a physical point of view,
 the influence of a finite correlation time  on  the phase diagram 
   is an  interesting  open  question~: does a finite
 correlation time favor or hinder  a noise-induced transition?  In particular,
    we wish to determine 
 how the shape of the  transition curve is modified when the noise is
 colored. In the  white noise case, 
 the asymptotic behavior of the critical curve when the amplitude
 of noise is either very small or very large is  known  explicitly
 and presents a simple  scaling  behavior 
 \cite{philkir1}. How do these scalings  change when the noise
 is correlated in time?       

 Due to the  finite  correlation time of the noise, the random oscillator 
  is  a  non-Markovian  
 random process and  there exists no  closed Fokker-Planck equation that 
 describes the dynamics of the Probability Distribution Function 
 (P.D.F.) in  the phase space. 
 This  non-Markovian feature hinders an exact solution in contrast with  the
 white noise case  where 
 a closed  formula for the  Lyapunov exponent was found.
   We shall therefore  have to 
 rely on various approximations to  carry out  an  analytical study
 of the phase diagram.  The 
 results obtained by  different approximations will be compared 
 with numerical results  and  with
 an  exact small noise perturbative expansion. 
 The various approximations have different regions of validity in the parameter
 space~: this  allows  us to derive  a fairly complete picture of the   
 phase diagram  of the random oscillator subject  to an  Ornstein-Uhlenbeck
 multiplicative noise.

 The  outline of this work is as follows. In section 2,  we 
  derive general results about the  stochastic
  oscillator with random frequency:
 thanks to dimensional analysis, we reduce the dimension of the 
 parameter space from four to two and  show how the Lyapunov exponent
 can be calculated by using an effective first order
 Langevin equation; we also 
 recall the  exact results  for white noise.  In section 3,
    we rederive  the rigorous 
 functional evolution  equation  of  P.D.F.;  although this equation is purely
 formal and  is not closed (it involves a hierarchy of correlation
 functions), it will be used as a systematic basis for various
   approximations; we  also
   carry out an exact perturbative expansion of the Lyapunov
   exponent in the small noise limit. 
  In section 4,  we consider a mean-field type approximation known as
  the `decoupling Ansatz' which provides  a simple expression
 for the colored noise Lyapunov exponent in terms of the white
  noise Lyapunov exponent. 
 In section 5, we consider two small correlation time approximations 
 that   both lead to an  effective Markovian evolution~: we show that
 these approximations are fairly accurate in the small noise regime. 
 In section 6,  we investigate the large correlation time
 limit  by   performing  an adiabatic elimination~:  this approximation
 is quite suitable for  the large noise regime.  
    The last section  is devoted to a synthesis and a discussion  
 of our results.

\section{General  results}
\label{sec:review}

\subsection{The random harmonic oscillator and the Lyapunov exponent}

  A  harmonic oscillator with a  randomly varying  frequency 
  can be described by the following equation 
\begin{equation}
\frac{\mathrm{d}^2 x}{\mathrm{d} t^2} + 
  \gamma \frac{\mathrm{d} x}{\mathrm{d} t}
  + ( \omega^2 + \xi_0(t)) x   = 0 \, ,
 \label{oscilrand1}
\end{equation}
  where $x(t)$ is the position of the oscillator at time $t$,
  $\gamma$ the (positive) friction coefficient and  $\omega$
  the mean value of the  frequency.  We assume 
  that the frequency fluctuations  $\xi(t)$  are modelised
 by  an Ornstein-Uhlenbeck
  process of amplitude ${\mathcal D}_0$  and of  correlation time
 $\tau_0$.  The long time behavior of  $x(t)$ 
 is characterized  by the Lyapunov 
 exponent defined as
 \begin{equation}
  \Lambda(\omega, \gamma, {\mathcal D}_0, \tau_0)  = \lim_{t \to \infty}
   \frac{1}{2t} \langle  \log   ( \frac{\dot x^2}{2}
    +  \frac{ x^2}{2} ) \rangle  = 
   \lim_{t \to \infty}
   \frac{1}{2t} \langle  \log E   \rangle   \, , 
\label{eq:deflyap}
\end{equation} 
where the brackets $\langle  \rangle$ indicate an averaging
 over realizations of the noise between 0  and $t$, {\it i.e.}, an averaging
  with respect to the Probability Distribution
 Function (P.D.F.) $P_t(x, \dot x)$;  
 the quantity $E$ is  the energy of the system.

  Taking  the time unit to  be  $\omega^{-1}$,
 we obtain  the following dimensionless parameters, 
\begin{equation}
  \alpha  =  \frac{\gamma}{\omega}   \,,  \,\,\,
 {\mathcal D}_1  = \frac{ {\mathcal D}_0 }{\omega^3} \,,  \,\,\,
  \tau_1 = \omega \tau_0 \, . 
\label{defalpha}
\end{equation}
 In terms of these
 parameters,  equation~(\ref{oscilrand1}) becomes
\begin{equation}
\frac{\mathrm{d}^2 x}{\mathrm{d} t^2} + 
  \alpha \frac{\mathrm{d} x}{\mathrm{d} t}
  + ( 1 + \xi(t)) x   = 0 \, .
 \label{oscilrand}
\end{equation}
 The  Ornstein-Uhlenbeck
 noise  $\xi(t)$  now has an  amplitude ${\mathcal D}_1$  and a 
   correlation time
 $\tau_1$   and   can  be  generated  from  the following  linear
  stochastic differential equation:
 \begin{equation} 
 \frac{{\textrm d} \xi(t)}{{\textrm d} t} = -\frac{1}{\tau_1} \xi(t) +
\frac{1}{\tau_1} \eta(t)  \, , 
  \label{OU}
\end{equation}
 $\eta(t)$ being  a Gaussian white noise
 of zero mean value and  of amplitude ${\mathcal D}_1$. In
 the stationary limit,  $\xi(t)$ has   exponentially
 decaying time correlations:
 \begin{eqnarray}
 \langle \xi(t) \xi(t') \rangle &=&  \frac{{\mathcal D}_1}{2 \tau_1}
   \exp\left( - |t - t'|/\tau_1    \right) \,   \, .  
 \label{statxi}
\end{eqnarray} 
  When $\tau_1 \to 0$, the process 
 $\xi(t)$ becomes identical to the white noise. 
 In terms of the dimensionless parameters, the Lyapunov exponent is 
 given by 
 \begin{equation}
  \Lambda(\omega, \gamma, {\mathcal D}_0, \tau_0)  =
 \omega  \Lambda(\alpha, {\mathcal D}_1, \tau_1)  \, .
\label{lyapscal1}
\end{equation}

\begin{figure}[th]
\centerline{\includegraphics*[width=0.40\textwidth, angle =-90]{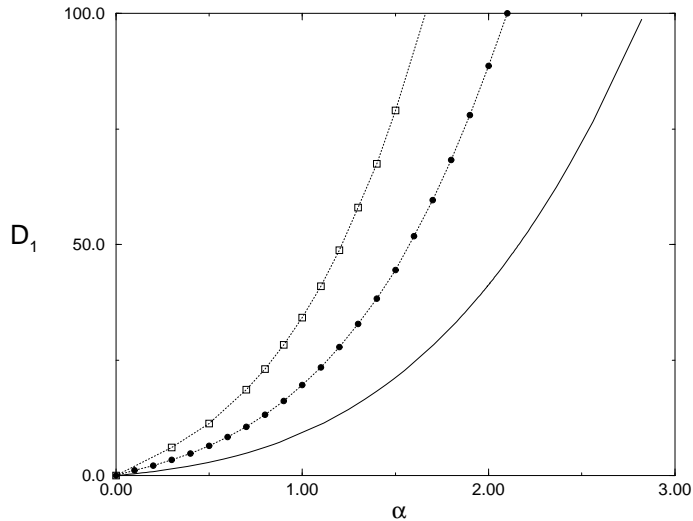}}
 \caption{\label{fig:simul} Critical curves  obtained
 by  simulating Eq.~(\ref{oscilrand}). 
 The curves with black circles  ($\bullet$)
 and with squares ($\Box$)  correspond to  $\tau =1$ and  
   $\tau =2$, respectively. The full black line is  the  analytical result
 for the  white noise case.   }
\label{figsimul}
\end{figure}

  The origin $(x = \dot x = 0)$ is a fixed point
  of equation~(\ref{oscilrand})  and in   the absence  of noise,
  it  is a stable and global attractor. If the  noise is 
    sufficiently strong,  the origin becomes unstable  and 
 the system  exhibits  an oscillatory behavior
   in   the stationary state.  Thus, the system can undergo   
  a noise induced  phase transition  
  from  an absorbing state to  a non-equilibrium steady state (NESS). 
 The Lyapunov exponent vanishes on the transition line between 
  the two phases~:  when $\Lambda < 0$,  the origin is stable
 and when    $\Lambda >  0$,  the stationary state   is  extended.
 In the extended phase, the system undergoes  random oscillations
  with increasing amplitude;  nonlinearities must therefore 
 be taken into account \cite{pmkmPRE, pmkmjstat}.
 In the vicinity of the phase transition line, on-off intermittency
 is displayed and the moments of $x$  
 exhibit a  multifractal scaling behavior \cite{aumaitre}. 
   Thus, for the random phase oscillator, the sign of  the Lyapunov  exponent
  determines the phase of the system.  
 In figure~\ref{figsimul}, we  plot this  transition 
 line computed  numerically
 for $\tau = 1.0 $ and 2.0. We  also draw   the
 transition line for the white noise, that was calculated
  analytically  in \cite{philkir1}.

The objective  of this  work is to study  various analytical  approximations 
 for the Lyapunov exponent when the noise is an Ornstein-Uhlenbeck process.
 These approximations will allow us to deduce  the  analytical features
 of  the phase diagram of a
 random frequency oscillator subject to an Ornstein-Uhlenbeck noise.

     \subsection{Reduced first order equation for the  Lyapunov exponent}

 As  in  the  white noise case,  the  Lyapunov exponent
 can be calculated by solving the following   first order nonlinear stochastic
 differential  equation
\begin{eqnarray}
   \dot z   &=&  f(z)  + \xi(t) \, ,
 \label{SDE1}   \\ 
  \mathrm{ with } \,\,\,   f(z)  &=& \epsilon -  z^2  \, ,
 \label{def:f}   \\ 
  \mathrm{ and  } \,\,\, \epsilon   &=&  \rm{sign}( \alpha -2)\, ;
 \label{def:epsilon}
\end{eqnarray}
the auxiliary variable $z$ is related to the original variable $x$
 by  $z =  \frac{\dot x}{x} + \frac{\alpha}{2}$. 
The  equation~(\ref{SDE1})  is derived from equation~(\ref{oscilrand})  in 
  Appendix \ref{dimensional}.  The  noise  $\xi(t)$   in this 
  equation  is an 
 Ornstein-Uhlenbeck process of amplitude $\Delta$
 and correlation time $\tau$,  {\it i.e.},  $\xi(t)$ satisfies 
 the relation 
  \begin{equation} 
     \dot   \xi  = -\frac{1}{\tau} \xi  +  
 \frac{1}{\tau} \eta   \, , 
  \label{OU1}
\end{equation}
where $\eta(t)$ is  a Gaussian white noise
of zero mean value and  of amplitude $\Delta$. 
  Thus, in the stationary limit we have
 $ \langle \xi(t) \xi(t') \rangle  =   \frac{\Delta}{2 \tau}
   \exp\left( - |t - t'|/\tau    \right) \,   \, .    $ 
 The parameters $\epsilon, \Delta$ and $\tau$ that appear
 in the reduced problem are related to  the dimensionless parameters
  $(\alpha, {\mathcal D}_1, \tau_1)$ 
 as follows (see  Appendix \ref{dimensional})   
  \begin{eqnarray}
    &\hbox{For }  \alpha <2 \hbox{ {\it(underdamped case)}}: 
  \,\,      &\epsilon = -1,   \,\,  \Delta  =  
 \frac{{\mathcal D}_1}
 {  \left( 1 - \frac{\alpha^2}{4}  \right)^{3/2}} \,, \,           
     \tau  =    \tau_1  \, \sqrt{1 - \frac{\alpha^2}{4} }
  \label{pmtrunder}  \, . \\
   &\hbox{For }   \alpha =2  \hbox{ {\it  (critical damping)}}: \,\,  
 &\epsilon = 0,   \,\, \Delta  =   1  \,,  \,             
     \tau  =  {\mathcal D}_1^{1/3} \tau_1  \,    \,,  \label{pmtcrit1}
  \\   &   \hbox{  or, equivalently, }   &\epsilon = 0, \,\, 
   \Delta ={\mathcal D}_1  \tau_1^3 \, , \,        \tau = 1  
 \label{pmtcrit2}   \, .  \\
  &\hbox{For }  \alpha  > 2 \hbox{ {\it  (overdamped case)}}:   \,\, 
  &\epsilon = +1,  \,\, \Delta  =  
 \frac{{\mathcal D}_1}
 { \left( \frac{\alpha^2}{4}-1  \right)^{3/2}} \, ,  
     \tau  =   \tau_1 \,  \sqrt{ \frac{\alpha^2}{4} -1 } \, .
 \label{pmtrover}
\end{eqnarray}

  We show in Appendix A that  the calculation of the   Lyapunov exponent 
 $ \Lambda( \alpha,  {\mathcal D}_1, \tau_1)$  of the random oscillator
 driven by noise can be reduced to the calculation of the 
following quantity~:
\begin{equation}
  \Lambda(\epsilon, \Delta, \tau)  =  
  \lim_{t \to \infty}\left( \langle  z   \rangle_t  
         + \frac{1}{2} \frac{d}{dt} 
   \left\langle \log\left( z ^2 +1 \right)\right\rangle_t
 \right)  \, .  
\label{deflambda2}  
\end{equation}
  Hereafter,  $\Lambda(\epsilon, \Delta, \tau)$ will be called
 the  {\it   reduced   Lyapunov exponent}. 
  Equation~(\ref{deflambda2})  seems to  involve 
 the time dependent P.D.F. of $z$, but  in Appendix B, we prove that 
\begin{equation}
  \Lambda(\epsilon, \Delta, \tau) = 
   \langle \,  z \, \rangle  \, , 
 \label{eq:idlyap}
\end{equation}
 where the stationary 
 average of $z$ is taken in the sense of principal parts. 
 The  formula~(\ref{eq:idlyap})
  requires  only  the knowledge  of the stationary P.D.F.
 of the random variable $z$ satisfying  the  
  Langevin equation~(\ref{SDE1}). 

The   Lyapunov exponent 
 $ \Lambda( \alpha,  {\mathcal D}_1, \tau_1)$  of the random oscillator
    is related to the   reduced    Lyapunov exponent
 $\Lambda(\epsilon, \Delta, \tau)$ as follows  (see  Appendix A)
\begin{eqnarray}
  \hbox{ For }  \alpha < 2, \, {\hskip 1cm  }
  \Lambda( \alpha,  {\mathcal D}_1, \tau_1) &=&  
     \sqrt{1 - \frac{\alpha^2}{4}} \,
   \Lambda(\epsilon = -1, \Delta, \tau)
     - \frac{\alpha}{2}     \, .   \label{lyapunder} \\
    \hbox{ For }   \alpha = 2,  \, {\hskip 1cm  }
   \Lambda(2,  {\mathcal D}_1, \tau_1) &=& 
   {\mathcal D}_1^{\frac{1}{3}} \,
  \Lambda(\epsilon = 0, 1, {\mathcal D}_1^{1/3} \tau_1)
     - 1   \, ,  \label{lyapcrit1}  \\  \hbox{ or, equivalently,}    \,\,\,\,\,\, 
 \Lambda(2,  {\mathcal D}_1, \tau_1)  &=&   \frac{1}{\tau_1}  
  \Lambda(\epsilon = 0, {\mathcal D}_1  \tau_1^3, 1) - 1 \, . 
   \label{lyapcrit2}  \\
    \hbox{ For }  \alpha > 2, \, {\hskip 1cm  } 
    \Lambda( \alpha,  {\mathcal D}_1, \tau_1) &=&    
     \sqrt{\frac{\alpha^2}{4}  -1} \, 
 \Lambda(\epsilon = 1, \Delta, \tau) 
     - \frac{\alpha}{2}    \, .   \label{lyapover}
\end{eqnarray}

 To summarize, the   mathematical  problem we have to solve  is to 
 find the stationary  P.D.F.  corresponding  to the
 Langevin equation~(\ref{SDE1})  and calculate  from it  the first moment
 of the  random  variable $z$. Then,  using 
 equations~(\ref{lyapscal1}),  ~(\ref{eq:idlyap})
 and~(\ref{lyapunder}, \ref{lyapcrit1} or \ref{lyapcrit2}, 
 \ref{lyapover}),  we can calculate  the Lyapunov exponent 
  in terms of the initial parameters of the
  stochastic oscillator.

     \subsection{The Lyapunov exponent for white noise}

 When the noise is white, the   Fokker-Planck equation for  $P_t(z)$, 
\begin{equation}
 \frac{ \partial P_t(z)} {\partial t} =  - 
  \frac{ \partial } {\partial z} \left( (\epsilon -z^2) P_t(z)\right) +
 \frac{\Delta}{ 2}
  \frac{ \partial^2 P_t(z)} {\partial z^2}   \, , 
  \label{FPwhite}
\end{equation} 
 can be solved exactly in the stationary limit \cite{philkir1}
 and we obtain
\begin{equation}
   P(z)   =  \frac{2J}{\Delta} 
\exp \left(\frac{2}{\Delta} (\epsilon z - \frac{z^3}{3}) \right)\,
 \int_{-\infty}^z {\rm d}u 
 \exp \left(-\frac{2}{\Delta} (\epsilon u -  \frac{u^3}{3}) \right) \, ,
 \label{pdfblanc}
 \end{equation}
 where the current $J$ is determined by the normalization condition
  $ \int_{-\infty}^{+\infty} {\rm d}z P(z)  = 1$.
 In \cite{philkir1}, we  calculated $\langle z \rangle$ and deduced the
 following expression for the  white noise Lyapunov exponent:
\begin{equation}
  \Lambda^{{\rm white}}(\alpha, {\mathcal D}_1)  = 
\frac{1}{2} \left\{  \frac {\int_0^{+\infty} 
\mathrm{d}u \; {\sqrt u} \;
  e^{ -\frac{2}{{\mathcal D}_1} \left(  ( 1 - \frac{\alpha^2}{4} ) u
   + \frac{u^3}{12} \right) }  }
 {\int_0^{+\infty}  \frac{\mathrm{d}u}{\sqrt u}
  e^{-\frac{2}{{\mathcal D}_1} \left(  ( 1 - \frac{\alpha^2}{4} ) u
   + \frac{u^3}{12} \right) } }
   -  \alpha \right\}   \, .       
\label{Lyapunovblanc}
\end{equation} 
 For small values  of the noise amplitude
 and  for  $\alpha < 2 $ the Lyapunov exponent  admits an expansion
 in powers of   ${\mathcal D}_1$  that begins  as follows:
   \begin{equation}
 \Lambda^{{\rm white}}(\alpha, {\mathcal D}_1) = \frac{  {\mathcal D}_1   }
 {8 ( 1 - \frac{\alpha^2}{4})}   - \frac{\alpha}{2} + 
 { \mathcal O}( {\mathcal D}_1^3) 
  =  \sqrt{1 - \frac{\alpha^2}{4}} \,   \frac{\Delta}{8}
   - \frac{\alpha}{2} + 
    { \mathcal O}( \Delta^3)   \, ,
\label{devblanc}
\end{equation}
where $\Delta$ is defined in equation~(\ref{pmtrunder}).
The stability boundary of the trivial solution
 $x = \dot x = 0$  is given by  
$\Lambda^{{\rm white}}(\alpha,{\mathcal D}_1) = 0$: this equation defines
 the critical line $ {\mathcal D}_1^c (\alpha) $
 in the parameter plane $(\alpha, {\mathcal D}_1)$. In \cite{philkir1},
 we obtained the implicit equation for this critical curve 
 and showed that  for small
 and large values of  $\alpha$, we have, respectively,
\begin{eqnarray}
  {\mathcal D}_1^c &\simeq& 4 \alpha        \nonumber    \\
  \hbox{ and  }  \,\,\, 
  {\mathcal D}_1^c &\simeq&  K  \alpha^3   \,\,\, \hbox{ with }
 \,\,\,  K \simeq  3.54 \, .
 \label{critblanc}
\end{eqnarray}
 At critical damping $\alpha = 2$, 
 the   expression of  the Lyapunov exponent
 is particularly simple   \cite{philkir1}
 \begin{equation}
 \Lambda^{{\rm white}}(2,  {\mathcal D}_1) = 
 \frac{\sqrt{\pi}}{\Gamma(1/6)}
 \left(\frac{3{\mathcal D}_1}{4} \right)^{\frac{1}{3}}     - 1  \,  .  
\label{blanc2}
\end{equation}

\section{Properties of the  colored noise P.D.F.}

  When the noise $\xi(t)$ is an Ornstein-Uhlenbeck process,
 the stationary  P.D.F. of the random variable
  $z$ driven by equation~(\ref{SDE1}) cannot be exactly determined.
  Thus  in contrast with  the white noise case, there   exists no closed 
  formula for the Lyapunov exponent $\Lambda$. In 
    the  first subsection,  we derive,  following
 \cite{hanggirev1},   the    exact functional evolution equation
 for  the P.D.F. with  colored noise. Although this equation 
 can not be solved, it will be useful  in the sequel
 to construct various  Fokker-Planck type approximations 
\cite{hanggirev1,sanchorev,JungHanggi, hanggirev2}. 
 In section \ref{sec:pert}, we derive 
 a  perturbative expansion  of $\Lambda$  in terms of  the 
 noise amplitude.  This expansion will be used in the following
 sections to test the accuracy of 
 some commonly used  approximations.

    \subsection{Functional  evolution equation for the  P.D.F.}

  When the noise  has  non-vanishing time
 correlations,  the dynamics of  $z$ is non-Markovian. 
 The evolution  of the  P.D.F. $ P_t(z)$  can no more  be described 
  by  a  closed Fokker-Planck  type equation.   The  P.D.F., rather,  
 evolves according to an integro-differential equation
 that involves a memory kernel, non-local in time. 
 Using a functional
 calculus approach, the   evolution equation of the P.D.F., defined as,  
\begin{equation}
  P_t(z) =  \langle \delta( z(t) -z) \rangle \, ,
 \label{def:PDF}
\end{equation} 
  can be derived  with the help of  the Furutsu-Novikov
  formula  and is given by 
\begin{equation}
 \frac{ \partial P_t(z)} {\partial t} =  - 
  \frac{ \partial } {\partial z} \left(  f(z) P_t(z)\right) +
 \frac{{\Delta}}{2 \tau}
  \frac{ \partial^2 } {\partial z^2} \int_0^t \rm{d}s \,\, 
         \exp\left( - |t - s|/\tau   \right) 
   \left\langle  \delta( z(t) -z)  \frac{\delta z(t)}{\delta \xi(s)}
 \right\rangle  \, ,
  \label{novikov}
\end{equation} 
where ${\delta z(t)}/{\delta \xi(s)}$ represents  the functional derivative of
the solution  $z(t)$  at time $t$  with respect to the value $\xi(s)$ of the
 stochastic process at time $s$. From  equation~(\ref{SDE1}), we find that
\begin{equation}
\frac{\delta z(t)}{\delta \xi(s)} = \theta(t-s)
 \exp\left(  \int_s^t f'(z(u)) \rm{d}u        \right) \, 
 \label{derivee}
\end{equation} 
 where $\theta(t-s)$ is the Heaviside function. Substituting this
 formula in   equation~(\ref{novikov}) leads to 
\begin{equation}
 \frac{ \partial P_t(z)} {\partial t} =  - 
  \frac{ \partial } {\partial z} \left(  f(z) P_t(z)\right) +
 \frac{{\Delta}}{2 \tau}
  \frac{ \partial^2 } {\partial z^2} \int_0^t \rm{d}s \,\, 
         \exp\left( - |t - s|/\tau   \right) 
   \left\langle  \delta( z(t) -z)  
  \exp\left(  \int_s^t f'(z(u)) \rm{d}u        \right) \,
 \right\rangle  \, . 
  \label{novikov2}
\end{equation} 
 Although   this evolution  equation for the  P.D.F.
   looks  like  a Fokker-Planck  equation, it   is not 
 a closed equation because it involves higher order correlations: 
 the functional derivative involves the knowledge of the function
 $z$ at different  times. It is only  in the case of white noise that 
 equation~(\ref{novikov2}) reduces  to the usual  Fokker-Planck  equation.
 Thus,  in order to make some progress, some closure assumptions
 must be made.

 \subsection{Small noise  perturbative expansion of the P.D.F.}
 \label{sec:pert}

  A  closed Fokker-Planck equation associated
  with  the Langevin equation~(\ref{SDE1}) can be constructed by
  a Markovian embedding of 
 the coupled stochastic equations~(\ref{SDE1} and \ref{OU1}).  
  However, we then have  to study a two variables Fokker-Planck
 equation for the joint P.D.F. $ P_t(z, \xi)$~: 
\begin{equation}
  \frac{ \partial P_t(z, \xi)} {\partial t} =
   - \frac{ \partial}{\partial z}
   \left(  (\epsilon - z^2 + \xi) P_t(z, \xi)    \right)
    + \frac{1}{\tau}\frac{ \partial P_t(z, \xi) }{\partial \xi}
     + \frac{\Delta}{2\tau^2}\frac{ \partial^2 P_t(z, \xi) }{\partial \xi^2}
  \, . 
 \label{eq:FPzxi}
\end{equation}
  This equation does not appear to be solvable  
 even in the stationary limit but it can be used to obtain 
  exact  perturbative  results \cite{crauel}.  We  derive 
  here  a second order 
  expansion of the Lyapunov exponent for  small
 noise and deduce from it the equation of the critical curve
 near the origin (both ${\mathcal D}_1$ and $\alpha$ are small)~:
  we  thus take  $\epsilon = -1$. We   write the  perturbative 
  expansion of the stationary 
  solution of  equation~(\ref{eq:FPzxi}) in  the form:
\begin{equation}
  P(z, \xi) = \sqrt{\frac{\tau}{\pi\Delta}}
 {\rm e}^{-\frac{\tau}{\Delta}\xi^2} \frac{1}{\pi(1 +z^2)}
  \Big(  A(z, \xi) + \Delta B(z, \xi) + {\mathcal O}(\Delta^2) \Big) \, ,
\label{eq:pert1}
\end{equation}
with
\begin{eqnarray}
   A(z, \xi)  &=&  1 + \xi a_1(z) +\xi^2 a_2(z) +\xi^3 a_3(z) + \ldots  \, ,\\
      B(z, \xi)&=&  b_0(z)  + \xi b_1(z) + \ldots  \, . 
\label{eq:pert1bis}
\end{eqnarray}
 The functions $a_i$ and $b_i$ satisfy a hierarchy of differential
 equations of first order in $z$ that can be solved recursively. 
The reduced Lyapunov exponent 
  $\Lambda(\epsilon = -1, \Delta, \tau)$ is 
 calculated from the following expression 
  (which has the advantage
  of  converging faster than  equation~(\ref{eq:idlyap}))
 \begin{equation}
  \Lambda (\epsilon = -1, \Delta, \tau) = 
  -  \left\langle  \frac{z \, \xi }{  1 + z^2}  \right\rangle 
\,, \label{formLambd}
\end{equation}
 where the average is calculated  with respect to   the stationary measure
 $P(z, \xi)$. Equation~(\ref{formLambd}) is 
 obtained by multiplying   equation~(\ref{eq:FPzxi})  by the factor 
 $1/2 \, \log\left( z ^2 +1 \right)$,   integrating  the right hand
 side by parts and taking  the limit $t \to \infty$. 
 Taking into account the fact that
 $\xi^2 \sim \Delta$ we observe  from equation~(\ref{formLambd})
  that  we must determine
 the functions  $a_1$, $a_2$, $a_3$, $b_0$ and $b_1$  if we want 
 to obtain   $\Lambda(\epsilon = -1, \Delta, \tau)$ up 
 to  the second order in $\Delta$. 
 After some lengthy calculations, we find 
 \begin{equation}
  \Lambda (\epsilon = -1, \Delta, \tau) =  \frac{ \Delta}{8( 1 + 4\tau^2)}  + 
  \frac{\Delta^2\, \tau \, 
 (624\tau^6 + 1632\tau^5 +176\tau^4 -312\tau^3 +137\tau^2 -324\tau +45)}
  {32 (1 +\tau^2)(1 +4\tau^2)^3(9 + 4\tau^2)}
                + {\mathcal O}(\Delta^3) \,.
\label{pertlambda}
\end{equation}

 This expansion  provides the local behavior of  the critical curve 
 near the origin. Using equations~(\ref{pmtrunder} and  \ref{lyapunder})
 and retaining only 
  the terms of the first order in $\alpha$,  we  find 
  the following equation for the critical curve ${\mathcal D}_1^c(\alpha)$~: 
\begin{equation}
   4(1 + 4\tau^2) \alpha   \simeq  {\mathcal D}_1^c 
  +  \frac{({\mathcal D}_1^c) ^2\, \tau \, 
 (624\tau^6 + 1632\tau^5 +176\tau^4 -312\tau^3 +137\tau^2 -324\tau +45)}
  {4 (1 +\tau^2)(1 +4\tau^2)^2(9 + 4\tau^2)}
 \label{critiquetaupetit}
\end{equation}

\section{The decoupling Ansatz}
\label{sec:decoupling}

\begin{figure}[th]
\centerline{\includegraphics*[width=0.40\textwidth, angle =-90]{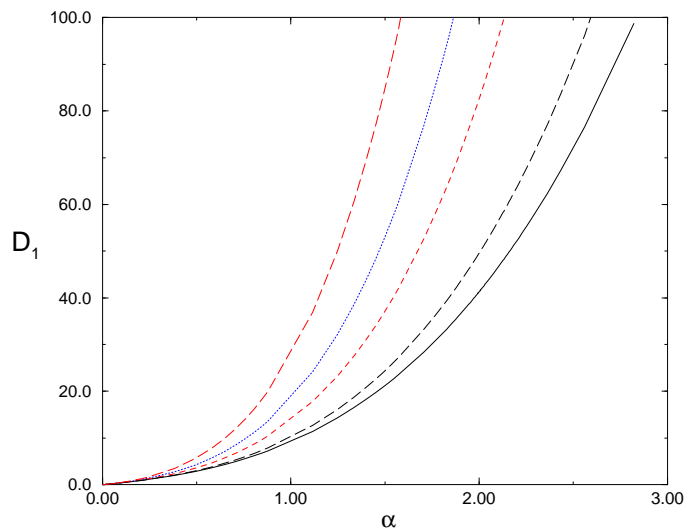}}
 \caption{\label{fig:decoupl} Phase diagram obtained
 by  the decoupling Ansatz for $\tau = 0.1, 0.5, 1.0$ and 2.0.
 The full black line corresponds to the white noise problem.}
\label{figdecoupling}
\end{figure}

 The decoupling Ansatz can be seen as a mean-field approach
 in which  higher order correlations are neglected. As such,
 it is  the simplest approximation from the technical point of view.
 It has the advantage that it does not  require any {\it a priori}
 assumption  on the correlation time of the noise.
 Starting from 
  equation~(\ref{novikov}), we  assume  that  the expectation
 value factorizes as follows~:
 \begin{equation}
 \left\langle  \delta( z(t) -z)  \frac{\delta z(t)}{\delta \xi(s)}
 \right\rangle  =    \left\langle    \delta( z(t) -z)  \right\rangle
   \left\langle   \frac{\delta z(t)}{\delta \xi(s)}  \right\rangle
     =  \theta(t-s) P_t(x)  \left\langle 
 \exp\left(  \int_s^t f'(z(u)) \rm{d}u \right)   \right\rangle       \, , 
\label{decoupl}
\end{equation}
 where the last equality  is obtained using equations~(\ref{def:PDF} and
 \ref{derivee}).  Using again a decoupling approximation and
 taking  the  stationary limit $ t, s  \to \infty$, we find
 \begin{equation}
\left\langle  \exp\left(  \int_s^t f'(z(u)) \rm{d}u \right)   \right\rangle
   =  \exp \left\langle  \int_s^t f'(z(u)) \rm{d}u \right\rangle
   = \exp\big( (t-s)  \langle    f' \rangle     \big)  \, ,
\end{equation}
 where the mean-value $\langle  f' \rangle $ is calculated in the
 stationary state. After inserting these approximations 
in equation~(\ref{novikov2}) and calculating
 explicitly the integral obtained, we find  the  effective Fokker-Planck
 equation for the decoupling approximation: 
\begin{equation}
 \frac{ \partial P_t(z)} {\partial t} =  - 
  \frac{ \partial } {\partial z} \left(  f(z) P_t(z)\right) +
 \frac{\Delta}{ 2(1 - \tau \langle  f' \rangle)}
  \frac{ \partial^2 P_t(z)} {\partial z^2}   \, .
  \label{EFPdec}
\end{equation} 
When $\tau \to 0$ this equation becomes identical to 
the white noise Fokker-Planck equation. As usual in mean-field
 approximations, the solution of equation~(\ref{EFPdec}) 
 is obtained by  solving the non-correlated case (i.e., the
 white noise problem) {\it supplemented with } a self-consistent 
 condition. Using  the expression~(\ref{def:f}) for 
 the function $f$, 
 we obtain the following  decoupled effective  Fokker-Planck equation
\begin{equation}
 \frac{ \partial P_t(z)} {\partial t} =  - 
  \frac{ \partial } {\partial z} \left( (\epsilon -z^2) P_t(z)\right) +
 \frac{\Delta}{ 2(1 + 2 \tau \langle  z \rangle)}
  \frac{ \partial^2 P_t(z)} {\partial z^2}   \, .
  \label{decoupled}
\end{equation} 
 This equation  is identical to  equation~(\ref{FPwhite})
  obtained for  white noise,  except for  the noise amplitude
 $\Delta$  that   is replaced here 
 by $\Delta/(1 + 2 \tau \langle  z \rangle)$. Recalling
 from equation~(\ref{eq:idlyap})   that 
$ \Lambda (\epsilon, \Delta, \tau)  =  \langle  z \rangle$,  we obtain 
 the following  relation 
\begin{equation}
 \Lambda (\epsilon, \Delta, \tau)  = 
  \Lambda^{{\rm white}} \left( 
 \frac{\Delta}{ 1 +
  2 \tau  \Lambda (\Delta, \tau)}   \right) \, .   
\label{implicite1}
\end{equation} 
 Reverting to the original parameters with the help
  of equations~(\ref{lyapunder}--\ref{lyapover}), we conclude that 
\begin{equation}
\Lambda( \alpha,  {\mathcal D}_1, \tau_1) = 
 \Lambda^{{\rm white}} \left(  \alpha, 
 \frac{ {\mathcal D}_1}{ 1 +  \alpha \tau_1 +
  2 \tau_1  \Lambda(\alpha,  {\mathcal D}_1, \tau_1) }   \right) \, , 
\label{implicite}
\end{equation}
  with    $\Lambda^{{\rm white}}$ 
  given by  equation~(\ref{Lyapunovblanc}). Equation~(\ref{implicite})
  is   an implicit expression for the Lyapunov exponent of
 the harmonic oscillator subject to Ornstein-Uhlenbeck noise 
 (this is an approximation because it is derived  from
 the decoupling Ansatz). The critical  line 
  ${\mathcal D}_1^c(\alpha)$ is obtained  by taking 
  $\Lambda(\alpha,  {\mathcal D}_1, \tau_1)  = 0$ and 
 its equation  is  
\begin{equation}
 0 =   \Lambda(\alpha,  {\mathcal D}_1, \tau_1) = 
 \Lambda^{{\rm white}} \left(  \alpha, 
 \frac{ {\mathcal D}_1}{ 1 +  \alpha \tau_1  }   \right) \, .
\label{lignetrans}
\end{equation} 
  The critical line for the Ornstein-Uhlenbeck
 noise, as obtained from the decoupling Ansatz, is thus 
 readily   deduced from the white noise  critical line  by a simple
 coordinate transformation. 
 The  resulting phase diagram 
 is drawn in figure~\ref{figdecoupling}; it agrees qualitatively
 with the numerically  computed  curves  
  given in figure~\ref{figsimul}.

  Using the expansions~(\ref{critblanc}) obtained 
  for the  white noise,
  we  derive  the asymptotic behavior of  the
  critical  line   in the decoupling approximation~:
\begin{eqnarray}
  {\mathcal D}_1^c &\simeq& 4 \alpha    \,\,\, \hbox{ for }  \alpha \ll 1 
   \label{asymptdec1} \\
   {\mathcal D}_1^c &\simeq&  3.54  \tau \alpha^4   
       \,\,\, \hbox{ for }  \alpha \gg 1  \label{asymptdec2}    \, .
\end{eqnarray}
 For small  values
  of  the dissipation rate $\alpha$, 
  the critical curve   obtained by   the decoupling approximation  is linear
 with the same slope  as  for  white noise~: 
  comparing  equation~(\ref{asymptdec1}) 
  with the exact expansion~(\ref{critiquetaupetit}), we observe
 that this result  is quantitatively correct
 only if  $\tau \ll 1$.   For large values
  of  the dissipation rate $\alpha$, 
 equation~(\ref{asymptdec2}) shows 
 that ${\mathcal D}_1$ scales as the fourth power of $\alpha$; 
  we recall that for white noise,  the asymptotic scaling
 is ${\mathcal D}_1^c \propto \alpha^3$ (see equation~(\ref{critblanc})). 
 This  asymptotic  scaling, 
  different  from the one  obtained in  the white noise case, is valid 
  as soon as $\tau \neq 0$, however small   $\tau$ may be.  
  This feature  will be investigated more specifically in section 6.

  Finally,   we can  study the special
 case $\alpha = 2$.  Defining  $L =  \tau_1
\Big( \Lambda(2,  {\mathcal D}_1, \tau_1)  + 1\Big) $, we find from 
  equations~(\ref{blanc2}) and ~(\ref{implicite})  that $L$ satisfies 
\begin{equation}
 L^3(1 +  2  L)     = \frac{ 3 \, \pi^{3/2}}{ 4 \, \Gamma(\frac{1}{6})^3}
   {\mathcal D}_1 \tau_1^3    \, . 
\label{eq:quartic}
\end{equation} 
 Solving  this  quartic  equation  for  $L$  leads to  
   the  Lyapunov exponent for the 
 critical value $\alpha =2$. In particular, we have 
\begin{eqnarray}
  \Lambda(2,  {\mathcal D}_1, \tau_1)  &\simeq&  \frac{\sqrt{\pi}}{\Gamma(1/6)}
 \left(\frac{3{\mathcal D}_1}{4} \right)^{\frac{1}{3}}     - 1    
 \simeq    0.289 \,\,  {\mathcal D}_1^{\frac{1}{3}}     - 1  
   \,\,\, \hbox{for } \,\,\,    {\mathcal D}_1 \tau_1^3 \ll   1 
   \label{asymptquartic1}       \, ,\\ 
  \Lambda(2,  {\mathcal D}_1, \tau_1)
   &\simeq&   \left(  \frac{\sqrt{\pi}}{\Gamma(1/6)}\right)^{\frac{3}{4}}
 \left( \frac{3 {\mathcal D}_1}{8 \tau_1}\right)^{\frac{1}{4}} 
        - 1  \simeq  0.331 \, 
   \left( \frac{{\mathcal D}_1}{ \tau_1}\right)^{\frac{1}{4}}   - 1
  \,\,\, 
 \hbox{ for } \,\,\,   {\mathcal D}_1 \tau_1^3 \gg  1 \, .
 \label{asymptquartic2}
\end{eqnarray}
 The  polynomial equation~(\ref{eq:quartic})  interpolates 
 between the  white noise solution
 (equation~(\ref{asymptquartic1}) is identical to   equation~(\ref{blanc2}))
  and  the  colored noise  scaling   with
 exponent  $1/4$  when $ {\mathcal D}_1 \tau_1^3 \gg  1$ 
 (equation~(\ref{asymptquartic2})).

\section{Effective  Fokker-Planck  equations for  small correlation time}
\label{sec:smalltau}

 In this section,  we study  the  effective Fokker-Planck equations 
  that are valid  when  the 
 correlation time $\tau$ of the noise is small.  
 The  stationary solutions of these equations  allow
 us to calculate small correlation time  approximations
  of   the Lyapunov exponent.  
 In the spirit of \cite{hanggirev1},  we derive  these approximations
  from the exact evolution equation~(\ref{novikov2}).

   \subsection{First order effective Fokker-Planck equation}

    In the small correlation time approximation, we assume that
 $\tau \ll 1$ and thus,   in equation~(\ref{novikov2}), we have $|t-s| \ll 1$.
 We then make the following approximation:
 \begin{equation}
\exp\left(  \int_s^t f'(z(u)) \rm{d}u \right)   \simeq
   \exp\big( (t-s)     f'(z(t))    \big)  \, .
\end{equation}
Inserting this expression in the functional equation~(\ref{novikov2})
 and using equation~(\ref{def:PDF}), we obtain an effective 
Fokker-Planck equation, valid at first order in $\tau$
\begin{equation}
 \frac{ \partial P_t(z)} {\partial t} =  - 
  \frac{ \partial } {\partial z} \left(  f(z) P_t(z)\right) +
 \frac{\Delta}{ 2}
  \frac{ \partial^2} {\partial z^2} 
  \Big( (1 + \tau  f'(z)) P_t(z)  \Big)     \, .
  \label{1stFP}
\end{equation} 
 We remark that in the limit $\tau \to 0$
  this equation reduces to the white noise
 Fokker-Planck equation. 
  Using  the expression~(\ref{def:f})  for  the
  function $f$,
 we find  that the first order effective Fokker-Planck
 equation leading to the Lyapunov exponent is given by 
\begin{equation}
 \frac{ \partial P_t(z)} {\partial t} =  - 
  \frac{ \partial } {\partial z} \left( (\epsilon -z^2) P_t(z)\right) +
 \frac{\Delta}{ 2}
  \frac{ \partial^2} {\partial z^2} 
  \Big( (1 -2 \tau z ) P_t(z)  \Big)     \, .
  \label{smallFP}
\end{equation}
We now solve this  effective Fokker-Planck equation in the stationary
 limit.  Introducing the stationary current $J$ we obtain 
\begin{equation}
   J  =  (z^2 - \epsilon)  P(z) +  \frac{\Delta}{ 2}
  \frac{ \partial} {\partial z} 
  \Big( (1 -2 \tau z ) P(z)  \Big)     \, .
  \label{statsmallFP}
\end{equation}
This equation is solved by the variations of constants method.
In terms of the  parameters
\begin{equation}
         z_0 = \frac{1}{2 \tau} \,\,\, \hbox{ and }  \,\,\,
          A = \frac{ 1 - 4 \tau^2 \epsilon}{4 \tau^3 \Delta} \, , 
\label{defA}
\end{equation}
the stationary P.D.F. can be written as 
\begin{eqnarray}
 \hbox{ For } \, z \le z_0, \,\,\, 
   P(z)  &=& \frac{2J}{\Delta} |2 \tau z -1|^{A-1} 
\exp \left(  \frac{z^2}{2 \tau \Delta} + \frac{z}{2 \tau^2 \Delta}  \right)\,
 \int_{-\infty}^z {\rm d}u 
\frac{\exp \left(-\frac{u^2}{2 \tau \Delta} - \frac{u}{2 \tau^2 \Delta}\right)}
{ |2 \tau u -1|^A}  \nonumber \\
 \hbox{  For } \, z \ge  z_0, \,\,\, 
   P(z)   &=& \frac{2J}{\Delta} (2 \tau z -1)^{A-1} 
\exp \left(  \frac{z^2}{2 \tau \Delta} + \frac{z}{2 \tau^2 \Delta}  \right)\,
 \int_z^{+\infty} {\rm d}u 
\frac{\exp \left(-\frac{u^2}{2 \tau \Delta} - \frac{u}{2 \tau^2 \Delta}\right)}
{ (2 \tau u -1)^A}  \, .
\label{PDFsmallFP}
\end{eqnarray}
 We remark that when $\tau \to 0$ this  P.D.F.
 becomes identical to the formula~(\ref{pdfblanc}) derived for white noise.
 Using   equation~(\ref{PDFsmallFP}), we can show that
 $ P(z)   \simeq \frac{J}{z^2} \,\,\, \hbox{ when } 
 \,\,\, z \to \pm \infty\,.$
  The behavior of  $P(z)$ when $ z  \to z_0$ is as follows:
\begin{eqnarray}
 &\hbox{  For }& \, A < 1 \, , \,\,\,  P(z)  \sim 
    \frac{1}{ |z -z_0|^{1-A}} 
 \, , \label{Ainf1}\\
 &\hbox{  For }& \, A = 1 \, ,  \,\,\, P(z)  \sim  \ln |z -z_0|
  \, , \label{A=1} \\
  &\hbox{  For }&  \,  A >  1   \, , \,\,\,  
 P(z_0) =  \frac{J}{\tau \Delta (A-1)}\, .  \label{Asup1}
\end{eqnarray} 
Thus, for $A > 0$, the stationary P.D.F. is a positive 
 and normalizable function and 
the current $J$ is fixed by imposing 
$\int_{-\infty}^{+\infty} P = 1$. The  small correlation
 time approximation breaks down when  $A < 0$,  
{\it i.e.},  when $ 4 \tau^2 \epsilon > 1$; this happens in the case 
  $\epsilon = +1$  and  $\tau > 1/2.$  In the rest of this section,
 we consider only the case $\tau <  1/2$~:  the  approximation is then  well
 defined   and  can be used to study  the Lyapunov exponent. 
 
  For  small noise, $\Delta \ll 1$  and low damping,
 $\epsilon = 1$, we   derive a perturbative expansion 
 of the stationary P.D.F.~(\ref{PDFsmallFP})~:   
\begin{equation}
  \Lambda(\epsilon = -1, \Delta, \tau)   =  \frac{ \Delta}{8}  + 
  \frac{\Delta^2\, \tau \,}{16}
                + {\mathcal O}(\Delta^3) \,.
\label{pertpetittau}
\end{equation}
 We observe that the  linear term in $\Delta$ agrees with the 
exact result~(\ref{pertlambda}) if terms of order 
  ${\mathcal O}(\tau^2)$ are neglected. However, the
 term proportional to  $\Delta^2$  does not coincide with the
  exact result even at the first order in
 $\tau$.  

  For  large noise,
 $\Delta \gg  1$,     and for  a given   correlation time $\tau$,  
     equation~(\ref{defA})  implies that  $A \to 0$.  From 
  equation~(\ref{Ainf1}), we conclude that   the stationary P.D.F.
 becomes more and more localized in the vicinity of $z_0$
 as $\Delta$ grows~:  hence
 $ \langle z \rangle \to  z_0 ={1}/{2\tau}$   as $\Delta \to \infty .$
  Thus, the  first order effective Fokker-Planck equation 
 predicts that  the Lyapunov exponent saturates 
  to a finite value when the amplitude 
 of the noise grows.  This prediction is unphysical and in contradiction
 with numerical results.  The first order effective Fokker-Planck equation
  is therefore meaningful only  for small amplitudes of the noise.

 In conclusion, 
 the first order  effective Fokker-Planck equation  does not provide
 a good approximation to calculate the Lyapunov exponent. 
 In the next section, we discuss a more sophisticated approach
 that leads to good  results  for  the small noise regime at least.

     \subsection{The `Best Fokker-Planck Equation'}

  This approximation has been first  proposed in \cite{lindcol1}. 
  Starting  from  an approximate integro-differential 
  equation for the P.D.F. valid at
 first order in the noise amplitude,  
 the evolution kernel  is calculated by a  resummation
   to all orders in the noise correlation time. This procedure results
 in an effective  Fokker-Planck equation, called the 
 Best Fokker-Planck Equation (B.F.P.E.). Although this equation 
 is not free from inconsistencies \cite{marchesoni}, 
 it  provides   in  some
 cases useful insights  that agree qualitatively with numerical simulations.
  (Another approach, that we shall not  follow  here, is to reject 
 the assumption of stationarity and to study non-conventional
 diffusion regimes of  a  Fokker-Planck equation
  with a time-dependent diffusion constant \cite{tsironis}.)

 Following  \cite{hanggirev1}, we   derive the B.F.P.E.  approximation   
  from the  exact functional evolution  equation~(\ref{novikov2}). In the
  B.F.P.E. approach the amplitude $\Delta$  of the  noise
  is supposed to be small and contributions  of  order $\Delta^n$
  with $n \ge 2$ are neglected. In order words, the exact evolution
 equation~(\ref{novikov2}) is replaced by 
   \begin{equation}
 \frac{ \partial P_t(z)} {\partial t} =  - 
  \frac{ \partial } {\partial z} \left(  f(z) P_t(z)\right) +
 \frac{{\Delta}}{2 \tau}
  \frac{ \partial^2 } {\partial z^2} \int_0^t \rm{d}s \,\, 
         \exp\left( - |t - s|/\tau   \right) 
   \exp\left(  \int_s^t f'({\bf \bar{  z}}(u)) \rm{d}u        \right) \,
   \left\langle  \delta( z(t) -z)  
 \right\rangle  \, ,
  \label{novBFPE}
\end{equation} 
where ${\bf\bar{z}}(u)$ is the  
solution of the  deterministic ( {\it i.e.,  noiseless})
 equation
 \begin{equation}
\frac{ {\rm d}{\bf\bar{z}}}{  {\rm d} u}
 = f\left(   {\bf \bar{z}}(u)        \right)  \,\,\, \hbox{ with }
  \,\,\,  {\bf \bar{  z}} (t) = z   \, .
\end{equation} 
 Substituting  this equation for  ${\bf\bar{z}}$ in 
   equation~(\ref{novBFPE}), we derive   the B.F.P.E.
  \begin{equation}
 \frac{ \partial P_t(z)} {\partial t} =  - 
  \frac{ \partial } {\partial z} \left(  f(z) P_t(z)\right) +
 \frac{{\Delta}}{2 } \frac{ \partial^2 } {\partial z^2}
 \Big(   D_t(z)  \,  P_t(z)  \Big)    \, \,\,\,  \hbox{ with  } \,\,\, 
 D_t(z)     =    f(z)   \int_0^t \rm{d}s \,\, 
     \frac{ \exp\left( - |t - s|/\tau   \right) }
 { \tau  f\left({\bf \bar{z}}(s)\right)}  \, . 
  \label{genBFPE}
\end{equation} 
The space and time dependent  diffusion factor  $ D_t(z)$ 
 can be explicitly  evaluated for  
 $f(z) = \epsilon - z^2$.
 In  the  long time limit  $t \to \infty$, we obtain   
 \cite{endnote}
 \begin{equation}
 \frac{ \partial P_t(z)} {\partial t} =  - 
  \frac{ \partial } {\partial z} \left(  f(z) P_t(z)\right) +
 \frac{{\Delta}}{2} \frac{ \partial^2 } {\partial z^2}
 \Big(    D(z)  P_t(z)  \Big) \, 
\,\,\,  \hbox{ with  } \,\,\, 
  D(z) = \frac{ 2\tau^2 z^2 - 2\tau z + 1 - 2 \tau^2 \epsilon }
  { 1 - 4 \tau^2\epsilon }  \, .
  \label{BFPE}
\end{equation} 
 The effective diffusion coefficient  $ D(z)$  is everywhere  positive  when
  $\epsilon = -1$ or  0 for all values of $\tau$.
 When  $\epsilon = 1$,  $ D(z)$ is everywhere 
  positive only if  $\tau <  1/2$. 
 In all these cases, we obtain 
 \begin{equation}
   P(z) =  \frac{2J}{\Delta}  
\exp \left( - 4\tau A z  \right)\,   D(z)^{-2A -1}
 \int_{-\infty}^z {\rm d}u \exp \left(  4\tau A u  \right)
    D(u)^{2A}       \, ,
  \label{PstatBFPE}
\end{equation} 
 where $A$ was defined in equation~(\ref{defA}). The value of  $J$
 is again fixed by  normalizing $P$. 
 Again,  when $\tau \to 0$,   this expression 
 becomes identical to the formula~(\ref{pdfblanc}) for the 
  P.D.F. obtained   in the white noise case.  From this expression of the
  stationary P.D.F. we derive the expression of the Lyapunov exponent:
\begin{equation}
  \Lambda(\epsilon , \Delta, \tau) =
   \frac{ \int_{-\infty}^{\infty} {\rm d}z
  \,  z \,  \exp \left( - 4\tau A z  \right)\,   D(z)^{-2A -1}
 \int_{-\infty}^z {\rm d}u \, \exp \left(  4\tau A u  \right)
    D(u)^{2A} }
{ \int_{-\infty}^{\infty} {\rm d}z 
   \exp \left( - 4\tau A z  \right)\,   D(z)^{-2A -1}
 \int_{-\infty}^z {\rm d}u \exp \left(  4\tau A u  \right)
    D(u)^{2A}  } \,, 
\label{LyapBFPE}
\end{equation}
 where the integral in the numerator  is defined  in the 
 sense of principal parts. This closed formula 
  can be used for a  numerical evaluation
 of the Lyapunov exponent  in the B.F.P.E. approximation
 (see figure \ref{figcompar}).

   The  perturbative expansion 
 of the Lyapunov exponent~(\ref{LyapBFPE})  when   $\Delta \ll 1$  and 
 $\epsilon = 1$  is given by    
\begin{equation}
  \Lambda(\epsilon = -1, \Delta, \tau) =  \frac{ \Delta}{8 ( 1 + 4\tau^2)}  + 
  \frac{\Delta^2\, \tau \, ( 1 + 3\tau^2) }{16( 1 + 4\tau^2)^2 }
                + {\mathcal O}(\Delta^3) \,.
\label{pertBFPE}
\end{equation}
 Comparing this expression with the exact result~(\ref{pertlambda}), 
 we observe that  the first term of this expansion
 with respect to $\Delta$ is correct. The B.F.P.E. 
 yields the exact  result  to all orders
 in $\tau$ and  performs  indeed a complete  resummation 
  with respect to the  correlation time. 
 However, the term  in  $\Delta^2$ does
 not agree with the exact result, even  in the $\tau \to 0$
 limit. We recall that  the BFPE approximation
 is intrinsically a small noise expansion and we see clearly, 
 in  this  specific example, that 
 this approximation is not valid
 at second order in the noise amplitude. This agrees with 
 the general analysis  of \cite{marchesoni} that 
  the BFPE can not be used for moderate values of the noise
  because the neglected  non-Fokker Planck terms 
  have a contribution of  the same order as the terms obtained after
 resummation with respect to the correlation time.

\section{Adiabatic  limit for  large correlation time}
\label{sec:taugrand}

In the previous section, we studied small correlation time
 effective equations which yield 
 fairly good approximations for the critical
 line for small-to-moderate values  of the noise amplitude.
  But   these approximations always 
  break down when $\alpha  \sim 1/\tau$:  
   a specific approximation scheme is  therefore   needed for this range
   and will become more and more relevant  as 
  $\tau$ grows. 

 In this last section, we thus study the large $\tau$ case.
  When  the correlation time is large,
 the Ornstein-Uhlenbeck noise becomes a slow variable and the
 dynamics of the random variable $z$  can be simplified. 
 We shall first perform an elementary  adiabatic elimination
  that yields an expression of the Lyapunov exponent in 
 the $\tau \to \infty $ limit. Then, we show that these results 
 can be obtained in a more  systematic manner thanks to the 
  Unified Colored Noise Approximation (UCNA).

\begin{figure}[th]
\centerline{\includegraphics*[width=0.40\textwidth, angle =-90]
{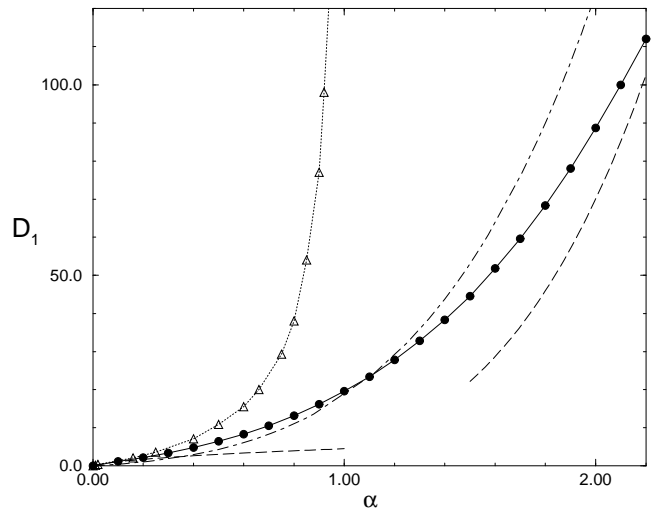}}
 \caption{ Various approximations
 for the phase digram at  $\tau =1$. The  critical curve, computed
 numerically,  is plotted
 in full line with  black circles ($\bullet$). The small   and large $\tau$
   expansions, equations~(\ref{pertlambda}) and (\ref{critadiab}) 
 respectively,
 are drawn with  dashed lines. 
 The decoupling approximation,  equation~(\ref{lignetrans}),
 is plotted with a  dot-dashed line. The BFPE approximation 
 is represented with $\triangle$.}
\label{figcompar}
\end{figure}

    When $\tau$ is large, the noise $\xi(t)$ is  slowly
 varying  and  we  make the approximation that  $z$  keeps
 adjusting  itself to   the  stationary   solution of  
 equation~(\ref{SDE1}). We  consider  here only  the
 domain  $\alpha > 2$ ({\it i.e.},  $\epsilon = 1$)
 for which the small correlation time expansions do not
 provide reliable results.  Thus, we study the limiting  case 
 where the noise  $\xi$ is quenched;  
 the equation~(\ref{SDE1}) then   reduces to
 \begin{equation} 
         \dot z  =   1  - z^2    + \xi \, ,
 \label{SDEadiab} 
 \end{equation}
    where $\xi$ is a {\it  time-independent}
  Gaussian random variable of variance $\Delta/(2\tau)$. 
 The value of  $z$ is then  given by  the stable fixed point
 of  equation~(\ref{SDEadiab}) if such a solution exists. We 
 must  distinguish  two cases: 
\begin{itemize}
 \item if $ \xi > - 1 $, then  $z  =   + \sqrt{ \xi + 1}$
 (the solution  $z  =   - \sqrt{ \xi + 1}$ is unstable);
  \item if $ \xi < - 1 $, then  equation~(\ref{SDEadiab})
 has no fixed point.  The random variable $z$  is given by
  $z(t) = A \, \mathrm{cotan} ( A t) $ with $ A^2=
 - (\xi  + 1)$. This running solution
  implies that  $z$  is  distributed over the whole  real axis.  
\end{itemize}
   From this discussion, we conclude that the distribution of $z$
 is given by
\begin{eqnarray}
   &\mathrm{For } \,\,   z > 0, &  
   P(z) = 2   \sqrt{ \frac{ \tau}{ \pi \Delta }} z 
       \exp\left( -\tau( z^2 - 1)^2/\Delta  \right)   
 +   \sqrt{ \frac{ \tau}{ \pi \Delta }}\int_{-\infty}^{-1}
 \frac{ \sqrt{ -\xi- 1)}}{\pi (z^2 - \xi - 1)}
           \exp\left( -\tau\xi^2/\Delta  \right)  \mathrm{d}\xi  \,, 
    \nonumber \\
   &\mathrm{for } \,\,   z < 0, &  
 P(z) =   \sqrt{ \frac{ \tau}{ \pi \Delta }}\int_{-\infty}^{-1}
 \frac{ \sqrt{ -\xi- 1)}}{\pi (z^2 - \xi - 1)}
           \exp\left( -\tau\xi^2/\Delta  \right)  \mathrm{d}\xi    \, .
  \label{distadiab}
\end{eqnarray} 
 This  P.D.F. consists of two parts: one term  is of Lorentzian type
 and  is an even  function of  z; the other term  exists only for $z >0$
 and is a rapidly decaying function. This contribution provides
 a strictly positive value for the mean value of $z$. Using
 equation~(\ref{distadiab}), we calculate the Lyapunov exponent
 \begin{equation}
   \Lambda(\epsilon = 1, \Delta, \tau) =
 \langle z \rangle =  2   \sqrt{ \frac{ \tau}{ \pi \Delta }}
     \int_0^{\infty}  z^2 
       \exp\left( -\tau( z^2 - 1)^2/\Delta  \right)   \mathrm{d} z
         \simeq     \frac{ \Gamma(3/4)}{2 \sqrt{\pi}}
          \Big(  \frac{\Delta}{\tau}  \Big)^{1/4}  \,\,\,\, \,\,\,\,
      \mathrm{ when }   \,\, \frac{\Delta}{\tau} \gg  1 \, .
\label{asymptadiab}
 \end{equation}
  This simple adiabatic approximation provides 
  the behavior of the Lyapunov exponent for large amplitude
  of the noise and it predicts a scaling in agreement with  
   the decoupling Ansatz
 (see equation~(\ref{asymptdec2})). Moreover,  it  explains 
 the existence of  long tails  for  the   P.D.F.  of $z$
 near infinity  ($P(z) \sim 1/z^2$ when $ |z| \to \infty)$
 and   the asymmetry  in $P(z)$ 
 for $z \to -z$.  Because of this asymmetry
 that favors  positive values of $z$,   the Lyapunov exponent,
 $\langle z \rangle$,   is always strictly positive.
 Using equation~(\ref{asymptadiab}), we obtain the
 asymptotic behavior of  the critical curve  for large values of the
 noise amplitude, 
  \begin{equation}
      {\mathcal D}_1^c \simeq 
 \left( \frac{ \alpha \sqrt{\pi}}{\Gamma(3/4)}   \right)^4 \tau_1
   \simeq 4.37 \, \alpha^4 \tau_1  \, . 
\label{critadiab}
 \end{equation}
 This approximation is  fairly  accurate  even  from  
 quantitative point of view and provides a good approximation
 of the critical noise amplitude even at moderate values: this can 
 be seen in  figure~\ref{figcompar},
 where the expression~(\ref{critadiab}) is compared with 
  the critical curve, computed numerically. 
 Moreover, the asymptotic behavior in equation~(\ref{critadiab})
  agrees well with the prediction of the decoupling
 Ansatz, equation~(\ref{asymptdec2}).

 For  the special case $\alpha =2$, we can repeat 
  the  above  calculations and  find that
\begin{equation}
\Lambda(2,  {\mathcal D}_1, \tau_1)  = \frac{ \Gamma(3/4)}{2 \sqrt{\pi}}
          \Big(  \frac{ {\mathcal D}_1}{\tau_1}  \Big)^{1/4}     - 1
   \simeq  0.345 \Big( \frac{ {\mathcal D}_1}{\tau_1}  \Big)^{1/4}  - 1  \, .
  \label{adiabal2}
\end{equation} 
  This expression is very similar  to   equation~(\ref{asymptquartic2})
   obtained by  decoupling Ansatz. 
   However the limit  $\tau \to 0$
 is ill-defined here; this is not a surprise because  the adiabatic
  elimination makes  sense  only if $\tau$ is large.

   Finally, we show that the   adiabatic elimination
  can be performed in a  systematic manner by using the simplest version
   of  the unified colored noise approximation 
  \cite{hanggirev1,JungHanggi,schimansky2}
  (for more sophisticated  UCNA schemes,
 see  {\it e.g.}, \cite{madureira,wio}). 
  Taking the time derivative of equation~(\ref{SDE1})
  and using  equation~(\ref{OU1}),    we obtain 
\begin{equation}
 \ddot z = f'(z)\dot z + \dot \xi =  f'(z)\dot z -\frac{1}{\tau} \xi
   + \frac{1}{\tau} \eta   \, .
\end{equation}
 Using again equation~(\ref{SDE1}) to express  the Ornstein-Uhlenbeck
  noise in terms of $z$ and the white noise  $\eta$,   we  derive   
    the following second-order
  nonlinear  Langevin equation driven by white noise:
 \begin{equation}
  \frac{ d^2 z}{dT^2} +   \gamma_\tau(z) 
   \frac{ d z}{dT} - f(z)   =   \frac{1}{\tau^{1/4}} \eta(T) \, 
  \,\,\,  \hbox{ where }  \,\,\, 
   \gamma_\tau(z) =  \frac{1}{\sqrt{\tau}} - \sqrt{\tau} f'(z)  \, ,
\end{equation}
where the new scale of time is   $T = t/\sqrt{\tau}$.
The effective damping coefficient  $\gamma_\tau(z)$ diverges
 when $\tau \to 0$ and $\tau \to \infty$. After an adiabatic elimination
 of the inertial term (valid for $ 2 \tau z + 1 \gg  \sqrt{\tau}$)
 the following effective stochastic equation
 is obtained:
 \begin{equation}
 \frac{ d z}{dT} = \frac{\epsilon  - z^2}{\gamma_\tau(z)} + 
  \frac{1}{\tau^{1/4}\gamma_\tau(z) } \eta(T)           \, ,
 \,\,\,\,   \hbox{ with } \, \,\,\, 
 \gamma_\tau(z) =  \frac{1}{\sqrt{\tau}} + 2  \sqrt{\tau} z \, .
\label{LangUCNA}
\end{equation}
  The solution of the  associated Fokker-Planck equation  leads
 to the stationary P.D.F., which in  
 the domain of validity of the UCNA  is   given by
\begin{equation}
   P(z)  =  {\mathcal N}  |2 \tau z +1|
\exp-\frac{2}{\Delta} \left( \frac{\tau z^4}{2} +
 \frac{z^3}{3} -\epsilon\tau  z^2 -   \epsilon  z \right)\,   \,  
 \hbox{ for } \, z \ge    -\frac{1}{2\tau}  \, .  
\label{PDFUCNA}
 \end{equation}
  The  constant ${\mathcal N}$ is 
 fixed by   the normalization condition on
 $P(z)$. 
 From this expression we deduce that the Lyapunov exponent
    is given by 
\begin{equation}
  \Lambda(\epsilon = 1, \Delta, \tau)   =  
   \frac{   \int_0^{\infty}  z |2 \tau z +1|
\exp-\frac{2}{\Delta} \left( \frac{\tau z^4}{2} +
 \frac{z^3}{3} - \epsilon\tau  z^2 - \epsilon  z \right) \mathrm{d} z }
 { \int_0^{\infty}  |2 \tau z +1|
\exp-\frac{2}{\Delta} \left( \frac{\tau z^4}{2} +
 \frac{z^3}{3} -  \epsilon \tau z^2 - \epsilon  z \right) \mathrm{d} z } \, .
\label{UCNAlambda}
\end{equation}
  Here, the range of the  variable  $z$ is taken  from 0 to $\infty$
  instead of $z \ge    -\frac{1}{2\tau}$ (this   is 
   a good approximation when $\tau$ is large). 
 For  $\epsilon = 1$ and 
  $ \Delta  \gg 1$, this  
  expression scales  as
  $({\Delta}/{\tau})^{1/4}$,   in agreement 
  with  the behavior predicted 
 by the decoupling Ansatz and  the adiabatic elimination. 
  For the special case $\alpha =2 $,
  equations~(\ref{lyapcrit2} and \ref{UCNAlambda})
  lead to 
\begin{eqnarray}
  \Lambda(2,  {\mathcal D}_1, \tau_1)  =  
\frac{1}{\tau_1}   \frac{   \int_0^{\infty}  z |2 z +1|
\exp-\frac{2}{ {\mathcal D}_1 \tau_1^3 } \left( \frac{ z^4}{2} +
 \frac{z^3}{3}\right) \mathrm{d} z }
 { \int_0^{\infty}  |2  z +1|
\exp-\frac{2}{{\mathcal D}_1 \tau_1^3 } \left( \frac{ z^4}{2} +
 \frac{z^3}{3} \right) \mathrm{d} z } - 1  
    &\simeq&    \frac{1}{\Gamma(2/3)}
 \left(\frac{3{\mathcal D}_1}{2} \right)^{\frac{1}{3}} - 1 
  \hbox{ for } \,\,\,  {\mathcal D}_1 \tau_1^3\ll 1    
  \, ,   \label{UCNAlalp2a}\\
    &\simeq&    \frac{\Gamma(3/4)}{\sqrt{\pi}}
 \Big( \frac{ {\mathcal D}_1}{\tau_1}  \Big)^{1/4} - 1 
             \hbox{ for } \,\,\,   {\mathcal D}_1 \tau_1^3 \gg  1 
\label{UCNAlalp2b}
\end{eqnarray}

 Thus,  the  UCNA approximation permits an  interpolation  between small
 and large  values of the correlation time. The predicted
 behavior matches the white noise scaling for  small $\tau$, 
 and the colored noise scaling at large $\tau$. 
  We remark that 
 prefactors in   equations~(\ref{UCNAlalp2a}  and \ref{UCNAlalp2b})  
 are different  from those   obtained in 
 equations~(\ref{blanc2}) and (\ref{asymptadiab}),
 respectively; this is due to the fact that
  in  the UCNA  the stationary
 current vanishes~: the running solution of $z$ are
  overlooked  and therefore  this approximation does not 
   describe satisfactorily  the tails of the P.D.F. of $z$.

 This study of the $\tau \to \infty$ limit shows
 that the 
 absorbing phase becomes  more and more stable as  
 the correlation time increases. Besides, it 
 confirms that 
 the existence of a  time correlation of 
 the noise renders the critical curve steeper   and modifies
 its scaling behavior for large values of the noise amplitude.

\section{Conclusion}

 The long time behavior  of the stochastic  oscillator with  random
 frequency is controlled by the sign of the Lyapunov exponent.
 In the case of a white noise perturbation, this  exponent
 can be calculated exactly and the phase diagram of the 
 stochastic  oscillator can  then  be rigorously  determined.  The aim of this
 work  is  to study the effects of time correlations on  the noise-induced
 bifurcation  of the stochastic  oscillator. In the case
 of an Ornstein-Uhlenbeck noise, an exact calculation  seems to be 
 out of reach and therefore we use  various approximation 
 schemes in order to derive analytical expressions for the 
 Lyapunov exponent and  to draw  the phase diagram.  Since  the 
 different approximations have distinct 
  regions of validity, their study 
 has allowed us to derive a  global picture 
 of the behavior of the system
 in the parameter space. In particular, we have derived  the scaling
 behavior  of the phase boundary in regions where the amplitude of the noise
 is small or large. Our results  agree fairly well with numerical
 simulations and  with exact perturbative expansions. These comparisons
 allow us to test the validity of  the different  approximation  schemes.

        We remark that the effective first order equation~(\ref{SDE1}),
   that we have used  to calculate
 the Lyapunov exponent, can be interpreted as describing
 an overdamped Brownian particle driven by colored noise
 in a metastable potential.  For such a system, a fundamental
  and extensively studied  quantity  is the mean escape-time
 of the particle \cite{vankampen} from the well.  The relation between 
 the  Lyapunov exponent and this mean escape-time deserves
 to be clarified. In particular, one could then use  path-integral  techniques
 \cite{bray,dykman} to  study  the phase diagram of the random
  frequency oscillator.  The  variation of
 the Lyapunov exponent  with  the correlation time of the noise could
  then be related to the phenomenon  of noise enhanced stability in metastable
 states \cite{dubkov}.

 The study carried out  here  for the  oscillator with
 multiplicative noise can be adapted to  other stochastic systems.
 For example,  Schimansky-Geier et al. \cite{schimansky} have shown 
 that  the  random Duffing  oscillator with {\it additive}  noise undergoes
 a subtle  phase transition that does not manifest itself
 in the stationary P.D.F. (which is simply given by the
 Gibbs-Boltzmann formula) but 
 affects the properties of the random attractor  
 in phase space. This  problem is   mathematically equivalent to 
  the noise-induced  bifurcation   of  a  linear oscillator subject
 to a multiplicative colored noise. This  noise  has
 a finite  correlation time and if we approximate it by an 
  Ornstein-Uhlenbeck process, the system becomes identical to the
 one studied here.  

  It has been suggested recently  \cite{vandenbroeck}
 that the Poisson process may be suitable  as a paradigm
  study of time correlation
 effects in random dynamical  systems.   In fact,  in many cases, 
  a complete analytical study can be carried out for a random variable
 driven by Poisson noise, but not when the stochastic force is  
  an  Ornstein-Uhlenbeck process.  We  have indeed 
  carried out the  exact calculation of the 
 Lyapunov exponent of a  random frequency oscillator subject to 
 a Poisson noise and obtained  the exact  phase diagram
 of the system \cite{orantin}. The results are in qualitative agreement
 with those found here with  an   Ornstein-Uhlenbeck  noise.

\appendix

\section{Derivation of  the effective Langevin equation}
\label{dimensional}

     In this appendix, we show that the calculation of the Lyapunov
 exponent can be reduced to solving a nonlinear  first order Langevin
 equation.   We introduce 
 the  function $y(t) =  \frac{\dot x}{x}$, that satisfies
\begin{equation}
\frac{\mathrm{d}  y}{\mathrm{d} t}  
  =   -(1 + \alpha y + y^2) +     \xi(t)      \, . 
 \label{oscilrand2}
\end{equation}

 Using the fact that
 \begin{equation}
 \frac{1}{2} \log(2E)  =  \log x + \frac{1}{2}  \log\left( y^2 +1 \right)\, ,
\end{equation}
 where $E = \frac{\dot x^2}{2}
    +  \frac{ x^2}{2} $ is the total  energy of the system, we derive
  the following identity:
\begin{equation}
 \frac{1}{2} \frac{d}{dt} \langle  \log E   \rangle 
       = \langle  y   \rangle_t  
         + \frac{1}{2} \frac{d}{dt} 
   \left\langle \log\left( y^2 +1 \right)\right\rangle_t  \, . 
\label{idapp}
\end{equation} 
This identity   implies that the Lyapunov exponent is given by 
\begin{equation}
  \Lambda (\alpha,  {\mathcal D}_1, \tau_1) =  \frac{1}{2}
  \lim_{t \to \infty}  \frac{d}{dt} \langle  \log E   \rangle 
  =  \lim_{t \to \infty} \left(\langle  y   \rangle_t 
         + \frac{1}{2} \frac{d}{dt} 
   \left\langle \log( y^2 +1)\right\rangle_t  \right)  \, . 
\label{eq:lyapapp}
\end{equation}

 We   introduce  an  auxiliary variable $z$ defined as
 \begin{equation}
    z =  \frac{\dot x}{x} + \frac{\alpha}{2} 
 =  y +   \frac{\alpha}{2}  \, .  
\label{eq:defz}
\end{equation} 
 Eliminating $y$  from  equation~(\ref{oscilrand2}),
 we obtain 
 \begin{equation}
    \dot z =   \frac{\alpha^2}{4} - 1  - z^2 + \xi(t)  \, .  
\label{eq:evolz0}
\end{equation} 
 We now show that  the dissipation parameter 
 can be  eliminated  from this equation   by
  a suitable redefinition of the parameters
 involved in the problem. We have to distinguish
 three cases: 
  \begin{itemize}
\item {\it Underdamped case} ($\alpha  < 2$):  in terms of 
   $ t :=   \sqrt{1 - \frac{\alpha^2}{4}}\,\, t$, the evolution equation
   of  $z$ becomes
 \begin{equation}
 \dot z =    - 1  - z^2 + \xi(t)  \, , 
 \label{eqy:under}
\end{equation}
where $\xi$ is an Ornstein-Uhlenbeck noise of amplitude
 and correlation time given  by 
\begin{equation} 
    \Delta  =  
 \frac{{\mathcal D}_1}
 {  \left( 1 - \frac{\alpha^2}{4}  \right)^{3/2}} \,, 
    \,\,\,\,\,\,    {\hskip  1  cm}            
     \tau  =    \tau_1  \, \sqrt{1 - \frac{\alpha^2}{4} }  \, .
 \label{Apmtrunder}
\end{equation}

\item {\it Critical damping}  ($\alpha = 2$):  the
 evolution of $z$ is given by 
\begin{equation}
 \dot z =     - z^2 + \xi(t)  \, . 
 \label{eqy:critical}
\end{equation}
 We  can rescale
 the time variable  either  as $ t := {\mathcal D}_1^{1/3}t$, or 
as $  t := t/\tau_1$. The  amplitude 
 and correlation time  of $\xi$  are then  given, respectively, by 
\begin{eqnarray}  
    &\Delta&  =   1  \,, 
    \,\,\,\,\,\,    {\hskip  1  cm}            
     \tau  =  {\mathcal D}_1^{1/3} \tau_1  \, .  \\ \hbox{ or } \,\,\,
   &\Delta& ={\mathcal D}_1 \tau_1^3  \label{Apmtcrit1}\,, 
    \,\,\,\,\,\,    {\hskip  1  cm}    \tau = 1 \, .
 \label{Apmtcrit2}
\end{eqnarray}
 We notice that 
  in  the critical damping case there is only one free  parameter 
 in the problem (in the white noise limit $\tau_1 =0$,  no free
   parameter is left). 

\item {\it Overdamped case}  ($\alpha >  2$): in terms of 
   $ t :=   \sqrt{\frac{\alpha^2}{4} -1} \,\, t$, the evolution equation
   of  $z$ becomes
\begin{equation}
 \dot z =    1  - z^2 + \xi(t)  \, ,  
 \label{eqy:over}
\end{equation}
where the  amplitude 
 and correlation time  of $\xi$ are    given  by 
\begin{equation}
     \Delta  =  
 \frac{{\mathcal D}_1}
 { \left( \frac{\alpha^2}{4}-1  \right)^{3/2}} \, , 
   \,\,\,\,\,\,  {\hskip  1  cm}  
     \tau  =   \tau_1 \,  \sqrt{ \frac{\alpha^2}{4} -1 } \, .
 \label{Apmtrover}
\end{equation}
\end{itemize}

  We have thus  shown  that the calculation of the Lyapunov exponent
of   the linear oscillator with random  frequency  can be reduced to 
 the study of the following equation 
 \begin{equation}
    \dot z =  \epsilon - z^2 + \xi(t)  \, ,   
\label{eq:evolz}
\end{equation} 
 with 
\begin{equation}
 \epsilon  =  \rm{sign}( \alpha -2) \, , 
\label{def:Aeps}
\end{equation} 
 {\it i.e.},  $\epsilon = -1, 0$ or 1 when 
   $\alpha <2$, $\alpha = 2$ and   $\alpha > 2$   respectively.
 Using equation~(\ref{eq:lyapapp}) and
  taking into account the  rescaling   of time, we conclude that 
  \begin{equation}
  \Lambda(\alpha,  {\mathcal D}_1, \tau_1)  =  \kappa 
    \Lambda(\epsilon,\Delta, \tau)    -  \frac{\alpha}{2} \, , 
\label{idapp2} 
\end{equation} 
 where we have defined
 \begin{equation}
  \Lambda(\epsilon, \Delta, \tau)  =  
  \lim_{t \to \infty}\left( \langle  z   \rangle_t  
         + \frac{1}{2} \frac{d}{dt} 
   \left\langle \log\left( (z - \frac{\alpha}{2})^2 +1 \right)\right\rangle_t
 \right)  \, , 
\label{deflambdaapp}
  \end{equation} 
 and  the coefficient $\kappa$  is equal to   
 $\sqrt{1 - \frac{\alpha^2}{4}}\,$,   ${\mathcal D}_1^{1/3}$
 or   $\sqrt{\frac{\alpha^2}{4} - 1 }\,$
  when   $\alpha <2$, $\alpha = 2$ or   $\alpha > 2$,    respectively.
 We remark that  when $t \to \infty$, 
\begin{equation} 
  \left\langle \log\left( (z - \frac{\alpha}{2})^2 +1 \right) \right\rangle_t
 -  \left\langle \log\left( z^2  +1 \right)\right\rangle_t
   \to    \left\langle  \log \frac{(z - \frac{\alpha}{2})^2 +1}
      { z^2  +1 }    \right\rangle   \, , 
 \end{equation} 
 where the  average  on the right hand side
  is taken with respect to the stationary P.D.F.
 of $z$  and does not depend on time. Therefore, equation~(\ref{deflambdaapp})
 is equivalent to 
\begin{equation}
  \Lambda(\epsilon, \Delta, \tau)  =  
  \lim_{t \to \infty}\left( \langle  z   \rangle_t  
         + \frac{1}{2} \frac{d}{dt} 
   \left\langle \log\left( z ^2 +1 \right)\right\rangle_t
 \right)  \, ,  
\label{deflambdaapp2}
  \end{equation}
   which is identical to equation~(\ref{deflambda2}).

\section{Proof of  equation~(\ref{eq:idlyap})}
\label{applyap}

 We  now derive the  identity~(\ref{eq:idlyap})  for the
 Lyapunov exponent that  involves only
 the stationary P.D.F. of $z$.  In all the cases considered
 in this work, we obtain  an effective Fokker-Planck equation  of the type~:
\begin{equation}
 \frac{ \partial P_t(z)} {\partial t} =  - 
  \frac{ \partial } {\partial z} \left(  f(z) P_t(z)\right) +
 \frac{\Delta}{ 2}
  \frac{ \partial^2} {\partial z^2} 
  \Big( D(z)  P_t(z)  \Big)     \, .
  \label{FPtype}
\end{equation} 
In the stationary limit, we have 
\begin{equation}
 J  = -  f(z) P(z) +  \frac{\Delta}{ 2} \frac{d} {d z} 
  \Big( D(z)  P(z)  \Big)   \, ,
  \label{FPstattype}
\end{equation} 
where $J$ represents the stationary current. 
  After an integration by parts,  we  deduce from 
 equation~(\ref{FPtype})  that
\begin{equation}
 \frac{d}{dt} 
   \left\langle \log( z^2 +1)\right\rangle_t  = 
  \int_{-\infty}^{+\infty} {\rm d}z \frac{2z}{ z^2 + 1} 
 \left( -  f(z) P_t(z) +  \frac{\Delta}{ 2} \frac{d(D(z)  P_t(z))} {d z} 
  \right)  \, .
\end{equation} 
Thus, we have
\begin{equation}
\langle  z   \rangle_t 
         + \frac{1}{2} \frac{d}{dt} 
   \left\langle \log( z^2 +1)\right\rangle_t =
     \int_{-\infty}^{+\infty} \,  z
      \Big(  P_t(z)
 -  \frac{ -  f(z) P_t(z) +  \frac{\Delta}{ 2} \frac{d}{dz}(D(z)  P_t(z)) }
 {z^2 + 1} \Big) {\rm d}z   \, . 
\end{equation} 
Taking the stationary limit  and using
  equation~(\ref{FPstattype}), we find 
\begin{equation}
 \lim_{t \to \infty}  \left( \langle  z   \rangle_t 
         + \frac{1}{2} \frac{d}{dt} 
   \left\langle \log( z^2 +1)\right\rangle_t \right) =
 \int_{-\infty}^{+\infty} \,  z \left(  P(z) - \frac{J}{z^2 +1} \right)
 {\rm d}z   \, . 
 \label{eq:subtract}
\end{equation} 
  Combining equations~(\ref{deflambdaapp2}) and (\ref{eq:subtract}),  
  we obtain  the following identity for the Lyapunov exponent
\begin{equation}
    \Lambda(\epsilon, \Delta, \tau) = 
\int_{-\infty}^{+\infty} \, z \left(P(z) - \frac{J}{z^2 +1} \right){\rm d}z  
 =    \int_{-\infty}^{+\infty} \,  z   P(z) {\rm d}z 
 \, ,  \label{lyapappend}
\end{equation} 
 where the last equality  has to be understood in the sense of
 calculating the 'principal part' of the integral, {\it i.e.}, 
\begin{equation}
  \int_{-\infty}^{+\infty} \,  z \,  P(z) {\rm d}z  = \lim_{M \to +\infty}
 \int_{-M}^{+M} \,  z \,  P(z) {\rm d}z \, .
\end{equation} 
Thanks to the formula~(\ref{lyapappend}), the Lyapunov exponent
is expressed in terms of  the stationary P.D.F. of $z$  
 and equation~(\ref{lyapappend}) can be rewritten as
\begin{equation}
  \Lambda =   \langle \,  z \, \rangle   \, , 
 \label{eq:idlyapB}
\end{equation} 
 which is identical to equation~(\ref{eq:idlyap}).
 We emphasize that the  first moment of the 
 stationary P.D.F. needs  to be defined  only in the sense of the
 principal values.


\begin{thebibliography}{article} 
\bibitem{vankampen} N.G. van Kampen,  {\it Stochastic Processes in Physics  
and Chemistry} (North-Holland, Amsterdam, 1992).
\bibitem{gardiner}  C. W. Gardiner, {\it Handbook of stochastic
  methods} (Springer-Verlag, Berlin, 1994).
\bibitem{anishchenko} V.S. Anishchenko, V.V. Astakhov, A.B.~Neiman, 
  T.E.~Vadivasova  and L.~Schimansky-Geier, {\it Nonlinear Dynamics 
 of Chaotic and Stochastic Systems} (Springer-Verlag, Berlin, 2002).
\bibitem{lefever} H. Horsthemke and R. Lefever,  {\it Noise
Induced Transitions}  (Springer-Verlag, Berlin, 1984).
\bibitem{fauve} R. Berthet,  S. Residori,
 B. Roman and  S. Fauve,  {\it Phys. Rev. Lett.}  {\bf 33}, 557 (2002); 
 F.~P\'etr\'elis, S.~Auma\^{\i}tre,
  {\it Eur. Phys. J. B}   {\bf 34} 281 (2003)
\bibitem{munoz}  M.  a. Mu\~noz,  {\it Nonequilibrium Phase transitions and
 Multiplicative Noise} in {\em  Advances in Condensed Matter
 and Statistical Mechanics}, edited by E. Korutcheva and R. Cuerno
 (Nova Science Publishers, 2004) {\it cond-mat/0303650}.
\bibitem{pikovskybook} A. Pikovsky, M. Rosenblum, and J. Kurths, 
 {\it  Synchronization,  A Universal Concept in Nonlinear Sciences}
  (Cambridge University Press, 2001).
\bibitem{luecke} M.~L\"ucke and F.~Schank,  {\it Phys. Rev. Lett.}
 {\bf 54}, 1465 (1985); M. L\"ucke, in  {\em Noise in Dynamical
 Systems, Vol. 2: Theory of Noise-induced Processes in Special
 Applications}, edited by F.~Moss and P.V.E.~Mc Clintock 
 (Cambridge University Press, Cambridge, 1989).
\bibitem{ebeling} W.~Ebeling, H.~Herzel, R.~Richert, L.~Schimansky-Geier,
  {\it Z. angew. Math. Mech.} {\bf 66} 141 (1986).
\bibitem{rong1}  H.W. Rong, G. Meng, X.D. Wang, W.~Xu and T.~Fang, 
   Journal  of  Sound and Vibration  {\bf 210}, 483 (1998).
\bibitem{rong2}  H. Rong, W. Xu and T. Fang, Journal  of  Sound and Vibration
 {\bf 283},  1250 (2005).
\bibitem{huang}  Z.~L.~Huang, W.~Q.~Zhu, Y.~Q.~Ni and J.~M.~Ko,
 Journal  of  Sound and Vibration  {\bf 254}, 245 (2002).
 \bibitem{xie} W.~C.~Xie,  Journal  of  Sound and Vibration 
   {\bf 289}, 171 (2006).
\bibitem{nayfeh}  A.~H.~Nayfeh,  {\em Perturbation Methods}
  (John Wiley, 1973);  A.~H.~Nayfeh, D.~T.~Mook, {\em Nonlinear Oscillations}
(John Wiley, 1979).
\bibitem{arnold} L.~Arnold, \emph{Random Dynamical Systems} 
(Springer-Verlag, Berlin, 1998).
\bibitem{pikovsky}R. Zillmer and A. Pikovsky, 
{\it Phys. Rev. E} {\bf 67} 061117 (2003).
\bibitem{bourret} R.C.~Bourret, {\it Physica}    {\bf 54},  623 (1971); 
          R.C.~Bourret, U.~Frisch and A.~Pouquet,
 {\it Physica}   {\bf 65}, 303 (1973).
\bibitem{philkir1} K. Mallick and P. Marcq,
{\it  Eur. Phys. J. B}  {\bf 36}, 119 (2003).
\bibitem{philkir2} K. Mallick and P. Marcq, {\it Eur. Phys. J. B}
  {\bf 38},  99    (2004). 
\bibitem{hansel} D.~Hansel and J.F.~Luciani, 
{\it J. Stat. Phys.}  {\bf 54}, 971 (1989).
\bibitem{tessieri} L.~Tessieri and F.M.~Izrailev, {\it Phys. Rev. E}
{\bf 62},  3090 (2000).
\bibitem{imkeller} P.~Imkeller and C.~Lederer, 
 {\it Dyn. and Stab. Syst.}  {\bf 14},  385 (1999).
\bibitem{pmkmPRE}    K. Mallick and P. Marcq, {\it Phys. Rev. E}
    {\bf 66}  041113   (2002).
\bibitem{pmkmjstat}  K. Mallick and P. Marcq,
 {\it J. Stat. Phys}  {\bf 119},  1-33 (2005).
\bibitem{aumaitre} S. Auma\^\i tre, F. P\'etr\'elis and K. Mallick,
   {\it Phys. Rev. Lett.} {\bf 95}  064101  (2005). 
\bibitem{hanggirev1}  P.~H\"anggi,   
in  {\em Noise in Dynamical Systems, Vol. 1},
edited by F.~Moss and P.V.E.~Mc Clintock (Cambridge University Press, 
Cambridge, 1989). 
\bibitem{sanchorev}  M.~San Miguel   and J.~M.~Sancho, 
 in  {\em Noise in Dynamical Systems, Vol. 1},
edited by F.~Moss and P.V.E.~Mc Clintock (Cambridge University Press, 
Cambridge, 1989). 
\bibitem{JungHanggi} P.~Jung  and  P.~H\"anggi,  
  {\it Phys. Rev. A}  {\bf 35}, 4464 (1987). 
 \bibitem{hanggirev2} P.~H\"anggi and P.~Jung, {\it  Adv. Chem. Phys.}
  {\bf 89}, 239 (1995).
\bibitem{crauel}   V. Wihstutz, in {\em Stochastic dynamics},
 edited by  H.~Crauel and M.~Gundlach (Springer Verlag, New-York, 1999). 
\bibitem{lindcol1} K.~Lindenberg and B.J.~West, 
{\it Physica A} {\bf 119}, 485 (1983);  K.~Lindenberg and B.J.~West, 
 {\it Physica A}  {\bf 128}, 25 (1984). 
\bibitem{marchesoni} P.~H\"anggi, F.~Marchesoni and P.~Grigolini,
{\it Z. Phys. B }  {\bf 56}, 333 (1984);    F.~Marchesoni, 
 {\it Phys. Rev. A} {\bf 36}, 4050  (1987).
 \bibitem{tsironis}  G.P. Tsironis and P. Grigolini,
   {\it Phys. Rev. Lett.}  {\bf 61}, 7 (1988). 
 \bibitem{endnote} We remark that the
  B.F.P.E.  can also be derived  using the original 
 operator formalism used in  \cite{lindcol1} 
 but we   preferred here to use a systematic approach 
 in which  the approximations are derived from the exact functional 
 equation~(\ref{novikov2}). 
% \bibitem{graham} R.~Graham, A.~Schenzle, Phys. Rev. A {\bf 26}, 1676 (1982).
 \bibitem{schimansky2} L.~H'walisz,  P.~Jung, P.~H\"anggi, P.~Talkner and 
L.~Schimansky-Geier, Z. Phys. B  {\bf 77}, 471 (1989).
\bibitem{madureira} A. J. R. Madureira, P. H\"anggi, V. Buonomano
   and W. A. Rodrigues, Jr.   {\it Phys. Rev. E} {\bf 51},  3849  (1995)
\bibitem{wio}  F. Castro, A. D. Sánchez  and H. S. Wio
   {\it Phys. Rev. Lett.}  {\bf 75},   1691  (1995);
  F.~Castro, H.~S.~Wio and G.~Abramson,  {\it Phys. Rev. E}
 {\bf 52}, 159  (1995).
 \bibitem{bray}  A.~J. Bray and  A.~J. Mckane, 
   {\it Phys. Rev. Lett.}  {\bf 62},   493  (1989).
 \bibitem{dykman}  M.~I. Dykman,  {\it Phys. Rev. A} {\bf 42}, 2020 (1990).
  \bibitem{dubkov}  A.~A. Dubkov, N.~V. Agudov  and B. Spagnolo,
   {\it Phys. Rev. E}  {\bf 69},  061103 (2004).
 \bibitem{schimansky}  L.~Schimansky-Geier and H. Herzel, {\it J. Stat. Phys.}
  {\bf 70}, 141  (1993).
  \bibitem{vandenbroeck}   I. Bena, C.~Van den Broeck, R.~Kawai
  and K.~Lindenberg, {\it Phys. Rev. E} {\bf 66} 045603(R) (2002);
 {\it Phys. Rev. E} {\bf 68},  041111 (2003); cond-mat/0501499.
 \bibitem{orantin} K.  Mallick and N. Orantin, {\it in preparation}.
\end{thebibliography}
\end{document}